\documentclass[dvipsnames,format=sigconf,anonymous=false,review=false,natbib=false]{acmart}
\AtBeginDocument{%
  }

\usepackage{amsmath, amsfonts,mathtools} 
\usepackage{units} 
\usepackage{eurosym}
\usepackage{csquotes} 
\usepackage[nocomma, short]{optidef}
\usepackage{color}
\usepackage{xcolor}

\copyrightyear{2024} 
\acmYear{2024} 
\setcopyright{rightsretained} 
\acmConference[GECCO '24 Companion]{Genetic and Evolutionary Computation Conference}{July 14--18, 2024}{Melbourne, VIC, Australia}
\acmBooktitle{Genetic and Evolutionary Computation Conference (GECCO '24 Companion), July 14--18, 2024, Melbourne, VIC, Australia}
\acmDOI{10.1145/3638530.3664160}
\acmISBN{979-8-4007-0495-6/24/07}

\RequirePackage[
  datamodel=acmdatamodel,
  style=acmnumeric,
  ]{biblatex}

\begin{document}

\title[Obtaining Robust Solutions From Quantum Algorithms]{Harnessing Inferior Solutions For Superior Outcomes: Obtaining Robust Solutions From Quantum Algorithms}

\author{Pascal Halffmann, Steve Lenk, Michael Trebing}
\authornote{All authors contributed equally to this research.}
\email{pascal.halffmann@itwm.fraunhofer.de}
\orcid{0000-0002-3462-4941}
\author{Michael Trebing}
\authornotemark[1]
\email{michael.trebingn@itwm.fraunhofer.de}
\affiliation{%
  \institution{Department of Financial Mathematics, Fraunhofer Institute for Industrial Mathematics ITWM}
  \streetaddress{Fraunhofer Platz 1}
  \city{Kaiserslautern}
  \country{Germany}
  \postcode{67663}
}

\author{Steve Lenk}
\affiliation{%
  \institution{Department of Cognitive Energy Systems, Fraunhofer IOSB}
  \streetaddress{Am Vogelherd 50}
  \city{Ilmenau}
  \country{Germany}}
\email{steve.lenk@iosb-ast.fraunhofer.de}

\renewcommand{\shortauthors}{Halffmann et al.}

\begin{abstract}
In the rapidly advancing domain of quantum optimization, the confluence of quantum algorithms such as Quantum Annealing (QA) and the Quantum Approximate Optimization Algorithm (QAOA) with robust optimization methodologies presents a cutting-edge frontier. Although it seems natural to apply quantum algorithms when facing uncertainty, this has barely been approached. 

In this paper we adapt the aforementioned quantum optimization techniques to tackle robust optimization problems. By leveraging the inherent stochasticity of quantum annealing and adjusting the parameters and evaluation functions within QAOA, we present two innovative methods for obtaining robust optimal solutions. These heuristics are applied on two use cases within the energy sector: the unit commitment problem, which is central to the scheduling of power plant operations, and the optimization of charging electric vehicles including electricity from photovoltaic to minimize costs.  These examples highlight not only the potential of quantum optimization methods to enhance decision-making in energy management but also the practical relevance of the young field of quantum computing in general. Through careful adaptation of quantum algorithms, we lay the foundation for exploring ways to achieve more reliable and efficient solutions in complex optimization scenarios that occur in the real-world.
\end{abstract}




\maketitle

\section{Introduction}\label{sec:intro}
Robust optimization problems \cite{BenTal2009RobustOptimization} are crucial in practical applications due to their ability to account for uncertain parameters inherent in real-world systems, making them indispensable for ensuring reliable and resilient decision-making. However, their complexity and the necessity to find a solution that copes well under multiple, often conflicting, uncertainty scenarios make these problems challenging to solve with conventional optimization techniques.

Quantum computing is heralded for its substantial computational power, which may be able to tackle computationally demanding problems more efficiently than classical computing, particularly in handling stochasticity and navigating complex search spaces. Its inherent characteristics are well-suited to exploring vast solution landscapes and managing the randomness central to robust optimization challenges. So far, only one approach has been postulated that tackles robust optimization with quantum computing. \textcite{Lim2023} present a quantum version of a well-known classical robust optimization algorithm characterized as online oracle-based meta-algorithm. This algorithm uses an oracle for the deterministic optimization problem in every iteration and the uncertain parameters are updating using online subgradient descent.

In this paper, we present a first glance at two methods for finding robust optimal solution given an optimization problem with uncertain parameters. The first method simply utilizes the fact that instead of providing one solution, a quantum algorithm such as QA \cite{Kadowaki1998} is executed multiple times, providing a set of samples. As (near-)optimal solution are more likely to occur and robust optimal solutions often are often near-optimal, we harvest the returned set of samples for the best solution given a robustness measure. The second algorithm is an adaption of the first method using QAOA \cite{Farhi2014}. Here, after we have found optimal parameters for QAOA using the expected value of the uncertain parameters, we run the QAOA circuit for each scenario and select the most robust solution.

The paper is organized as follows. In the remainder of this section, we briefly introduce robust optimization and formally describe the two use cases we are considering, which are the unit commitment problem \cite{Halffmann2023,koretsky2021adapting} dealing with finding an optimal production schedule for power units subject to satisfying energy demand and energy production constraints and the problem of finding a cost-optimal charging schedule while utilizing energy coming from photovoltaic power. In Section \ref{sec:solharvest}, we introduce the quantum annealing approach and in Section \ref{sec:2StepMethod}, we present the adjustments made for QAOA. We show an example for both approaches. We close with an outlook on further research in the last section. 

\subsection{Overview on Robust Optimization}\label{sec:robust}
Robust optimization deals with optimizing decisions under uncertainty (e.g., uncertain data, model parameters) by finding solutions that are feasible under a range of possible scenarios and are rather immune to variations in the parameters of the optimization problem. There are two central robustness concepts:

\textbf{Min-max worst-case robustness} focuses on finding the solution that performs best in the worst-case scenario. This concept is formulated as $\min_{x \in X} \max_{\xi \in \Xi} f(x, \xi)$, where \(x\) represents the decision variables within the feasible set \(X\), \(\xi\) represents uncertain parameters within their uncertainty set \(\Xi\), and \(f(x, \xi)\) represents the objective function that is to be minimized.

\textbf{Min-max regret robustness} aims to minimize the worst-case regret, which is the difference between the outcome of a decision and the best outcome that could have been achieved with perfect information. It is formulated as $\min_{x \in X} \max_{\xi \in \Xi} \left( f(x, \xi) - f^*(\xi) \right)$, where \(f^*(\xi) = \min_{x' \in X} f(x', \xi)\) represents the best possible outcome for a given scenario \(\xi\).
Remark that we have not specified the uncertainty set \(\Xi\) containing the different scenarios \(\xi\). In general, this set can be a discrete set, an interval for a certain parameter of the optimization problem, or even a more complex set. For our use cases the demands, given for each time step, are identified as uncertain and we have a discrete set of demand scenarios that can occur. For those interested in delving deeper into robust optimization, we refer to \textcite{BenTal2009RobustOptimization} who provide a comprehensive overview and an excellent starting point. Further, \textcite{Gabrel2014RecentAdvancesInRobustOptimization} review developments in and applications of robust optimization.

\subsection{Overview on the Use Cases}\label{sec:usecases}
The UCP is tasked with finding a cost-minimal operation schedule for a set of thermal units $i\in\{1,\ldots,N\}$ to meet a given demand for electricity over a distinct set of time steps $t\in\{1,\ldots T\}$. Further, technical properties of both the thermal units and the underlying power grid have to be respected. Take note that there does not exist one single UCP, but several variants are present in the literature, see \textcite{Knueven2020} for an overview. In our UCP, each thermal unit $i$ has the following properties: linear, production dependent costs $varcost_i$ and fixed costs $startcost_i$ for starting the unit, minimum and maximum power generation output, $mingen_i$, $maxgen_i$, and minimum running time and minimum idle time, $minup_i$, $mindown_i$. At each time step $t$ the residual demand $rd_t$, demand minus supply by renewable energies plus spinning reserve, has to be met. Limitations due to the power grid are omitted, giving the following optimization problem:

\scriptsize
\begin{mini*}[2]
	{}{\sum_{t=1}^T \sum_{i=1}^I \left(varcost_i\cdot gen_{t,i} + startcost_i\cdot start_{t,i}\right) }{}{}
	\addConstraint{\sum_i gen_{t,i}}{=rd_t,}{\forall~t=1,\ldots,T,}
	\addConstraint{on_{t,i}\cdot mingen_i}{\leq gen_{t,i},}{\forall~t=1,\ldots,T,~i=1,\ldots,I,}
	\addConstraint{on_{t,i}\cdot maxgen_i}{\geq gen_{t,i},}{\forall~t=1,\ldots,T,~i=1,\ldots,I}
	\addConstraint{on_{t,i}-on_{t-1,i}}{\leq start_{t,i},}{\forall~t=1,\ldots,T,~i=1,\ldots,I,}
	\addConstraint{\sum_{\tau=t}^{t-1+minup_i}on_{\tau,i}}{\geq start_{t,i}\cdot minup_i,}{\forall~t=1,\ldots,T,~i=1,\ldots,I,}
	\addConstraint{\sum_{\tau=t+1-mindown_i}^{t}start_{\tau,i}}{\leq 1-on_{t-mindown_i,i},}{\forall~t=1,\ldots,T,~i=1,\ldots,I,}
	\addConstraint{on_{t,i}, start_{t,i}}{\in\mathbb{B}}{\forall~t=1,\ldots,T,~i=1,\ldots,I,}
	\addConstraint{gen_{t,i}}{\in\mathbb{R}_{\geq 0}}{\forall~t=1,\ldots,T,~i=1,\ldots,I.}
\end{mini*}
\normalsize
We have three sets of decision variables $on_{t,i}$, $gen_{t,i}$ and $start_{t,i}$, $t\in\{1,\ldots T\}, i\in\{1,\ldots,N\}$. The binary variable $on_{t,i}$ observes whether unit $i$ is running at time $t$. The variables $start_{t,i}\in [0,1]$ track the starting of power units. Further, we have used a logarithmic encoding of the power generation output by using binary variables $gen_{t,i,k}$ for each time step $t$, each power unit $i$. For a certain stepsize $step_i$ for each power unit $i$ we have $k=1,\ldots,d_i\coloneqq \lfloor\log_2\left(\nicefrac{(maxgen_i-mingen_i)}{step_i}\right)\rfloor + 1$.
Since quantum algorithms for optimization problems, often require the optimization problem written as an Ising Hamiltonian, which has a close resemblance to a \emph{Quadratic Unconstrained Optimization Problem (QUBO)}, we utilize the QUBO formulation as presented in \cite{Halffmann2023}.

Analogously to \cite{Federer2023GI,Federer2022VPPC, Federer2022QCE, Federer2022GI}, we describe the problem of minimizing the costs of EV fleet charging: The goal is to minimize grid usage costs and maximize the local PV consumption, which earns less money per kW/h than grid usage costs.  However, since day-ahead forecasts for the solar power generation are rather inaccurate, uncertainties occur through the standard deviation $\sigma_t$ in mean solar power $\mu_t$. The PV power generation $E_t^{pd}$ implicitly includes the local energy demand. The maximal allowed charging power $j_{t}$ is included to represent the number of EVs to be charged. By trivial assumptions the maximal charging power per EV is set to $1$. Charging power is restricted to be in the limits of $j_t^{min}$ and $j_t^{max}$ and the overall charging is between $E^{min}$ and $E^{max}$. We get the following optimization problem:

\scriptsize
\begin{mini*}[2]
	{}{C(j_t)\coloneqq\mathbb{E}\left[\sum_t (j_t - E_t^{pd})^2\right]}{}{}
	\addConstraint{j_t^{min}\leq j_t}{\leq j_t^{max},}{\forall~t=1,\ldots,T,}
    \addConstraint{E^{min}\leq \sum_{t=1}^T j_t}{\leq E^{max},}{}
	\addConstraint{j_t}{\in\mathbb{B}}{\forall~t=1,\ldots,T.}
\end{mini*}
\normalsize
Here, the expectation value $\mathbb{E}[.]$ is taken over the stochastic variations of the problem instances, i.e.\ the PV system variations. By assuming a Gaussian distributed PV energy $E_t^{pd}$ the expectation value can be calculated explicitly. Together by assuming a uniform distribution of the PV power between the limits $E_t^{pd,lo}$ and $E_t^{pd,up}$, we can rewrite the problem as stated in \cite{Federer2022VPPC} as follows:
\scriptsize
\begin{equation*}
    \min_j \sum_t \left( j_t^2 - C_t j_t + D_t\right)\, 
\end{equation*}
\normalsize
with appropriately chosen coefficients $C_t$ and $D_t$. From this form we can see that stochastic variations in the presented forms do not imply an increased complexity in comparison to the deterministic approach above. However, in practice, a two-step approach is often used to consider uncertainties in optimization, in which the expectation value of the uncertainty is used in the first step and in the second the probability of the approach or respective scenarios are included. Including the probability distribution would require an analytical derivation of the target function, which is often not easily achievable. Thus, scenario approaches employing  Monte Carlo simulations to select respective scenarios are most often used. Classically, this approach is computationally intense since it requires the solution of the optimization program for each scenario and a postprocessing step to find the stochastic solution.

\section{Solution Harvesting for QA}\label{sec:solharvest}
QA is a quantum algorithm for solving QUBO problems by finding the state of lowest energy, the ground state, of a Hamiltonian $H_P$, an operator corresponding to the total energy of a quantum mechanical system. This ground state is equivalent to  the minimal solution of a QUBO. The approach of QA is based on adiabatic evolution: starting from a simpler Hamiltonian $H_I$ already in its ground state, we look at the time dependant system $H(t)=t\cdot H_P + (T-t)\cdot H_I$ for $t\in[0,T]$. Now, if we move slowly through time, the system remains at the ground state and at time $T$, we obtain  the ground state of $H_P$ in the sample set with high probability, given a number of shots of QA.

Now, our approach is fairly straightforward: the sample set returned by QA does not only contain the optimal solution but also inferior solutions, in particular near-optimal solutions. Thus, instead of writing a robust optimization problem as a QUBO, which would require many qubits and may not be solvable with today's quantum hardware, we use the corresponding deterministic problem with the expected values of the uncertain parameters as input and search the stochastic output for the most robust solution given a robustness concept. Since robust optimal solutions are not far off of the optimal solutions for the individual scenarios, it is likely that the sample set of QA also contains a robust (near-)optimal solution. This is a two step approach: In the first step, we build the QUBO corresponding to the deterministic problem with expectation values of the uncertain parameters, and we apply QA obtaining a set of samples. Next, in the postprocessing of QA, we rank the solutions by their robustness, either their worst-case robustness value or regret robustness value. The best robust solution is then returned.

\begin{figure}[tb]
    \centering
    \includegraphics[width=0.45\textwidth]{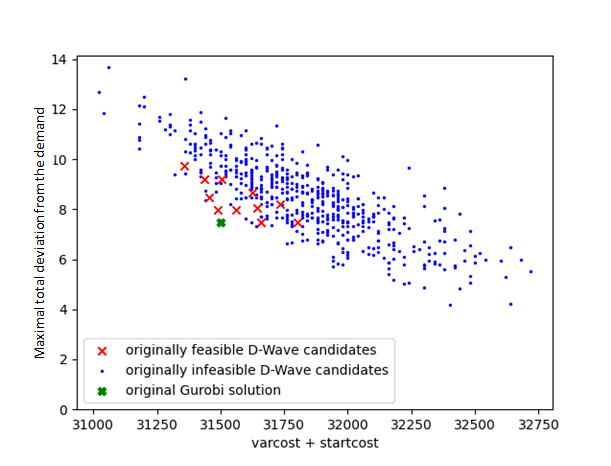}
    \caption{The results of the experiment with the deterministic part of the objective function value on the x-axis and the regret regarding the demand on the y-axis.}
    \label{fig:testrobust}
\end{figure}

We apply this approach to the UCP of Section \ref{sec:usecases}, where the residual demand (demand minus supply of renewable energies) is uncertain. Thus, given a set of scenarios for the residual demand in each time step, we use the expected residual demand in each time step for the deterministic QUBO. Remark that it can occur that the produced amount of energy does not satisfy the demand, therefore we utilize the objective function value of the QUBO instead of the linear program  Nevertheless, we filter out solutions that do not satisfy the power unit properties such as minimum running and idle time. For testing, we apply this approach using the D-Wave Advantage System JUPSI in Jülich, Germany, and compare it with solving the min-max regret robust optimization problem via Gurobi. We test this on an instance with 2 power plants and 12 time steps and 25 scenarios for the demand, see Figure \ref{fig:testrobust}. Clearly, while we are not able to find the robust optimal solution, the feasible solutions obtained with D-Wave are close to the robust optimal solution returned by Gurobi. Thus, our approach shows potential as an effective heuristic for robust optimization problems and should be tested further with suitable metrics.

\section{QAOA-based Robust Optimization}\label{sec:2StepMethod}
We adapt the approach of Section~\ref{sec:solharvest} to QAOA and provide a new variant. QAOA is a hybrid-classical algorithm described as an variational algorithm. This iteratively executes a parameterized quantum circuit and optimizes the parameters using the measurement of the circuits by a classical optimization routine. It is closely connected to QA mimicking the traversing from the initial to the problem Hamiltonian by using two blocks of parameterized circuits, one for the problem Hamiltonian and one so-called mixer. In contrast to QA, QAOA provides us with the parameters and the building blocks with more control and variability.

Similar to QA, we employ QAOA on the problem with expected values of the uncertain parameters and obtain optimal QAOA parameters, e.g., using a grid search-based method \cite{Federer2022VPPC}. We deviate now from the previous approach: Instead of harvesting the output of QAOA for the best robust solution, we adjust the problem Hamiltonian and thus the QAOA parameterized circuit slightly for each scenario.
We execute this circuit using the parameters obtained in the first step. For each scenario we obtain a sample set and search for the most robust solution in all sample sets, given a robustness concept. This approach includes the set of the scenarios and not just the expected value of the uncertain parameters. Further, it is also possible to be applied in the realm of stochastic optimization: If probabilities of the scenarios are known or these are themselves sampled from a probability distribution, the probability can be used to determine the sample size for each scenario. Then, the samples and their frequency can be used to calculate stochastic measures like expectation value or expected value of perfect information \cite{Shapiro2009}.

\begin{figure}[tb]
    \centering
    \includegraphics[width=0.35\textwidth]{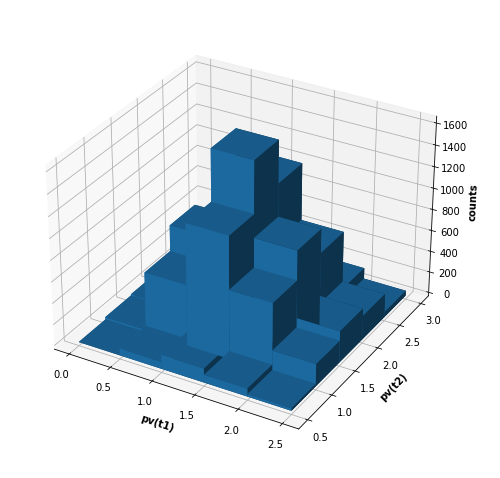}
    \caption{Distribution of the scenarios of the charging scheduling example for $25$ scenarios given as a 3D histogram. Each scenario consists of an energy supply for time step 1 ($pv(t1)$ and time step 2 ($pv(t2)$).}
    \label{fig:Scens}
\end{figure}

\begin{figure}[tb]
    \centering
    \includegraphics[width=0.35\textwidth]{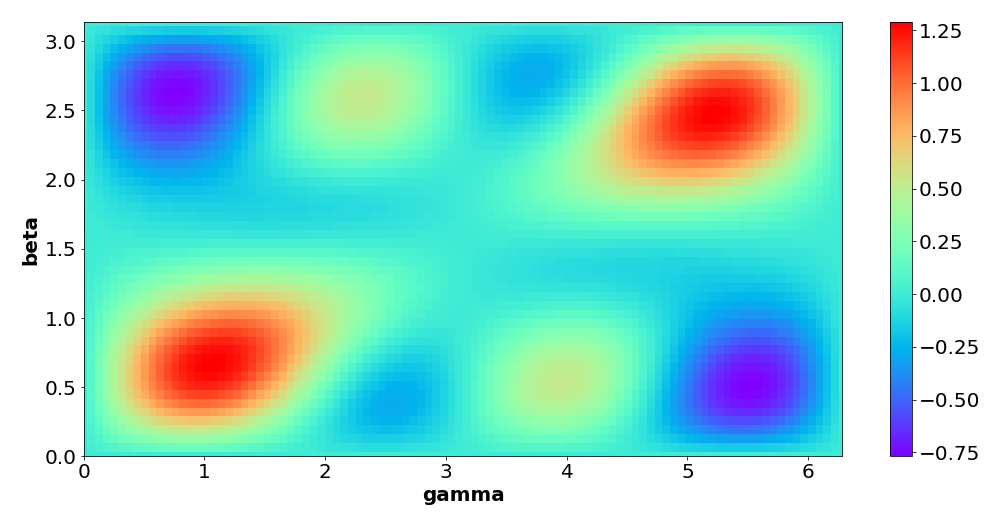}
    \caption{Energy expectation value landscape for our example. We use the expectation value of the PV energy supply $\mu$ for the determination of the optimal variational parameters $\beta$ and $\gamma$ of the QAOA.}
    \label{fig:ExpLand}
\end{figure}

This method is applied to the EV charging scheduling problem of Section.~\ref{sec:usecases} with uncertain PV energy supply. We are considering a problem with two time steps and 25 scenarios for the PV supply. The histogram of the 25 scenarios is given in Figure~\ref{fig:Scens}. We execute the circuits using a quantum simulator and no real hardware. For the first step, Figure \ref{fig:ExpLand} shows the corresponding energy landscape of QAOA we traverse in our search for the optimal parameters for QAOA. After executing QAOA for each scenario (100 shots per scenario), we are considering the min-max regret robustness here as well. In our small experiment, we obtain the robust-optimal charging schedule with a min-max regret of 1.9 that is also obtained via Gurobi. Similar to Section~\ref{sec:solharvest}, this motivates further investigations of the performance of our approach.

\section{Outlook}
The exemplary results from both experiments demonstrate promising potential for leveraging quantum computing to effectively solve robust optimization problems. Further investigations including multi layer QAOA and a well-defined full benchmark with a set of test instances may verify this hypothesis. Additionally, if the probability is included in the QAOA approach, quantum computing can be used to solve stochastic optimization problems with different stochastic measurements as well. We will pursue this idea in the future. Additionally, a sensitivity analysis is ongoing to evaluate the limits of the proposed approaches due to the variations of the expectation value landscape for the scenarios.

\begin{acks}
We acknowledge funding by Bundesministerium für Wirtschaft und Energie, Germany through the project “EnerQuant” (Grant No.~03EI1025A) and gratefully acknowledge the Jülich Supercomputing Centre (https://www.fzjuelich.de/ias/jsc) for providing computing time on the D-Wave Advantage™ System JUPSI through the Jülich UNified Infrastructure for Quantum computing (JUNIQ) project “Robust optimization for the Unit Commitment Problem” (Project No.~28042). P.H. was additionally funded by Bundesministerium für Bildung und Forschung (BMBF), Germany through the project “QuSAA” (Grant No.~13N15695) and S.L. by the European Union under Horizon Europe Programme - Grant Agreement 101080086 — NeQST. Views and opinions expressed are however those of the author(s) only and do not necessarily reflect those of the European Union or European Climate, Infrastructure and Environment Executive Agency (CINEA). Neither EU nor granting authority can be held responsible for them.
\end{acks}

\printbibliography


\end{document}